\documentclass[conference]{IEEEtran}
\IEEEoverridecommandlockouts
\usepackage{cite}
\usepackage{amsmath,amssymb,amsfonts}
\usepackage{amsthm}
\usepackage{algorithmic}
\usepackage{graphicx}
\usepackage{textcomp}
\usepackage{xcolor}
\newtheorem{definition}{Definition}
\usepackage{tikz}
\def\BibTeX{{\rm B\kern-.05em{\sc i\kern-.025em b}\kern-.08em
    T\kern-.1667em\lower.7ex\hbox{E}\kern-.125emX}}
\usepackage{pgfplots}
\pgfplotsset{compat=1.18}
\usepackage{pgfplotstable}
\usepackage{siunitx}
\usepackage{subcaption}
\usepackage{pgfplots}

\begin{document}

\title{Decoding OTC Government Bond Market Liquidity: An ABM Model for Market Dynamics \\ 
\thanks{This work is funded in part by an ARC Laureate grant FL200100204}
}

\author{\IEEEauthorblockN{1\textsuperscript{st} Alicia Vidler}
\IEEEauthorblockA{\textit{dept. Engineering} \\
\textit{UNSW, Australia}\\
a.vidler@unsw.edu.au}
\and
\IEEEauthorblockN{2\textsuperscript{nd} Toby Walsh}
\IEEEauthorblockA{\textit{dept.Engineering} \\
\textit{UNSW, Australia}\\
t.walsh@unsw.edu.au}
}

\maketitle

\IEEEoverridecommandlockouts	\IEEEpubid{	\parbox{\columnwidth}{\vspace{+ 5 \baselineskip} 
~\copyright 2024 IEEE. Personal use of this material is permitted.  Permission from IEEE must be obtained for all other uses, in any current or future media, including reprinting/republishing this material for advertising or promotional purposes, creating new collective works, for resale or redistribution to servers or lists, or reuse of any copyrighted component of this work in other works. Dec 15, 2024 \hfill}	\hspace{0.3\columnsep}\makebox[\columnwidth]{\hfill}}	\IEEEpubidadjcol

\begin{abstract}
The over-the-counter (OTC) government bond markets are characterised by their bilateral trading structures, which pose unique challenges to understanding and ensuring market stability and liquidity.  In this paper, we develop a bespoke ABM that simulates market-maker interactions within a stylised government bond market.  The model focuses on the dynamics of liquidity and stability in the secondary trading of government bonds, particularly in concentrated markets like those found in Australia and the UK. Through this simulation, we test key hypotheses around improving market stability, focusing on the effects of agent diversity, business costs, and client base size. We demonstrate that greater agent diversity enhances market liquidity and that reducing the costs of market-making can improve overall market stability. The model offers insights into computational finance by simulating trading without price transparency, highlighting how micro-structural elements can affect macro-level market outcomes. This research contributes to the evolving field of computational finance by employing computational intelligence techniques to better understand the fundamental mechanics of government bond markets, providing actionable insights for both academics and practitioners.

\end{abstract}


\section{Introduction}

The liquidity of OTC government bond markets plays a critical role in the functioning of modern financial systems, influencing both market stability and economic policy implementation. Unlike equity markets, which are centralised and provide large amounts of public transaction data, OTC bond markets operate through decentralised networks of participants, with comparatively little public data, often creating challenges in understanding market dynamics and liquidity flows. Liquidity, in this context, refers to the ease with which bonds can be bought or sold within a reasonable time frame, or rather \textit{"liquidity is the ability to trade when you want to trade"} \cite{Harris2002TradingAE}.  Despite its importance, market liquidity in OTC bond markets remains difficult to quantify and predict, particularly during periods of stress when liquidity can evaporate suddenly, triggering broader financial instability \cite{Pinter2023}. 



To address this, we propose an ABM designed to simulate the micro-level interactions of key participants (clients and market makers) within the OTC government bond market. ABM's are particularly suited to this challenge, as they allow the examination of complex, decentralised systems where individual behaviours and local interactions give rise to emergent phenomena at the macro scale \cite{Wooldridge_2009}. A particular challenge for researches of bilateral bond markets is that data in this field is sparse and bond trade level data is not freely available \cite{Pinter2023}. To address this we model the dynamics between market makers, the data for which is partially available to the public, isolating their trading behaviours and liquidity provision through agent to agent trading and client interaction. In addition to liquidity, we are also concerned with the related concept of market stability. The World Bank defines financial market stability as \textit{"the absence of system-wide episodes in which the financial system fails to function"} \cite{WorkBank24}, with the International Monetary Fund going so far as to define the stability of certain financial markets, such as that of the UK system as a "global public good" \cite{FSIF20203}. 

In this study, we first design an ABM that is able to replicate observable Australian market characteristics. We select the Australian government bond OTC market to design our bespoke ABM, leverage its unique blend of international relevance and local specificity. Whilst smaller than the UK, or Canada, the Australian market shares essential structural features with larger ones due to their common heritage of British market design. Like the UK, the Australian market is characterised by a small number of key market makers (MM), operating in a decentralised, over-the-counter regulated framework. The Australian market offers a relative richness and accessibility of data. Detailed reports from the Australian Office of Financial Management (AOFM) and other government bodies provide a public picture of market activity, such as aggregated MM market share and market structure. This comparative level of transparency allows for stylised calibration of ABMs, making the Australian bond market an attractive choice for academic study.  There is no reason to suppose that results formed on the Australian market cannot be generalised to other markets (e.g. UK, Canada). 


In this paper, we investigate key aspects of market liquidity and stability. First, we examine how the diversity of market makers influences these factors. Our simulations reveal that increasing the heterogeneity of market makers—through varying trading costs and client base sizes—enhances market trading, liquidity, and stability more effectively than merely increasing their number. Diverse agent characteristics promote more frequent interactions and stabilise liquidity provision. In contrast, markets populated by homogeneous agents, regardless of their quantity, tend to become fragile and unstable after a certain number of trading steps. This underscores the importance of fostering diversity in trading behaviours and strategies to maintain a resilient market structure.

Our second key hypothesis explores the impact of reducing operational costs for agents and expanding their client bases. We find that lowering business costs for market makers leads to a more stable and liquid market by boosting transaction volumes and extending the duration of stable operations. These findings highlight the nuanced relationship between agent characteristics, market design, and liquidity, offering insights into regulatory policies that could enhance market efficiency.

The paper is organised as follows. In Section 2, we review the literature on ABM design and bond market modelling. Section 3 introduces our custom ABM, detailing how we represent key market features within a framework inspired by Axtell's Sugarscape model. Section 4 presents our results. Finally, Section 5 concludes with a brief summary and discusses areas for future research \footnote{Code available :https://github.com/AliVidl/ABM-for-Bond-Markets }.


\section{Literature Review: Bond Markets and Agents}

Agent-based models lack a consensus approach for calibration \cite{avegliano2019using}. The bilateral nature of bond market trades hinders direct observation of agents' utility functions and policies \cite{boggess2022toward}, impeding the application of machine learning techniques successfully used in calibrating ABMs for observable markets like foreign exchange \cite{Ardon2021, vadori2022multiagent, ganesh2019reinforcement}. These techniques often rely on features absent in bilateral bond markets, such as homogeneous agent classes, observable environments, or comprehensive regulatory data \cite{Banks2021, barzykin2021algorithmic, cont2021stochastic}.


Global bond markets, including the Australian government bond market, involve diverse participants acting as agents: banks as trading facilitators (market makers), institutional investors (debt purchasers), and trading firms like hedge funds (clients) \cite{BOE}. Unlike centralized equity exchanges, bond trading occurs over-the-counter (OTC) across fragmented electronic platforms. The Australian bond market exemplifies this decentralised landscape, lacking a dominant trading venue and operating through opaque bilateral transactions facilitated by the four major retail banks designated as market makers \cite{BondMarket2023}, while being systemically important to both Australia and the global trading landscape \cite{RBA}.



Government bond prices are significantly impacted by central banks' interest rate policies, inflation expectations \cite{Duffie_1999}, and periodic interest payments, unlike stock markets driven by supply and demand. Although bond prices inversely relate to market-expected yields and can price in potential rate changes, they largely reflect current rate policies—publicly available information. Holding a bond provides interest returns, making access to bond market liquidity more important, as returns can be generated by holding the asset. Our contribution is to model the movement of bonds across market participants, focusing on liquidity flow characteristics.

\subsection{Market Micro-structure and Trading Venues}  
The lack of publicly available data for empirical validation is a key challenge in simulating bilateral markets like bond markets \cite{Pinter2023}. While macro trading data indicates domestic market makers traded between themselves on average 27.23\% of all secondary market bond turn over in Australia since 2016\footnote{https://www.aofm.gov.au/data-hub}, granular details on individual trades and participant strategies remain obscured. Bilateral transactions are mediated through a handful of electronic messaging platforms \cite{BondMarket2023}. Recent efforts to launch a joint venture centralising bond market transaction data collection \cite{TradeWebJointVenture2023} underscore possibilities for future research. In line with regulatory demands for transaction visibility \cite{Duffie2020}, our work offers a fresh modelling approach to the evolving market micro structure. Due to the current lack of data, our model relies on stylised facts and government documentation. 
 


\subsection{Bond trading modelling}  
Literature on bond markets predominantly addresses bond pricing models \cite{Longstaff_1995, Duffie_1999, Black_1976, Merton_1974, Liang_2017} and alternative pricing methods \cite{AlbaneseVidler}, with limited exploration of market design aspects \cite{Braun-Munzinger2018, SEC, Lin_2020}. Prior to 2008, bond markets grappled with interest rate volatility and uncertainty, making bond price models crucial. Recent work using ABMs to simulate interactions between interdependent agents \cite{Bai_2020, Fagiolo_2017, Lussange_2020, Vermeir_2015} discusses challenges in capturing the complex dynamics across various applications. Integrating concepts of the Adaptive Market Hypothesis \cite{Lo_2004}, ABM frameworks like SugarScape \cite{Axtell} and MESA \cite{Pike_2019, Masad_2015, Pike_2021, Kazil_2020} have been applied to study artificial societies and complex systems. Our contribution combines and expands these areas to bond trading and liquidity provision in OTC markets without price transparency.

Recent advancements include incorporating reinforcement learning techniques to train agent models in financial markets, often reducing participants to two homogeneous groups: liquidity providers or consumers \cite{vadori2022multiagent, BROCK19981235}. Our work moves beyond this binary categorisation to include heterogeneous agents for increased model richness.

\subsection{Crisis and Market Structure Research}
Studies examining market structure and liquidity in the UK government bond market \cite{Pinter2023} provide a detailed analysis of the 2022 UK gilt market crisis, underscoring the fragility of government bond markets and inspiring our modelling. The work highlights risks posed by concentrated market makers and sudden liquidity evaporation, paralleling crises in other sovereign bond markets, including U.S. Treasury market disruptions in 2020 \cite{Duffie2020, Treasury2011}. These events reveal vulnerabilities in OTC markets where liquidity providers struggled during extreme volatility, emphasising the critical role of regulatory frameworks and market design in maintaining stability, particularly in markets dominated by a few key participants. A major contribution of our model to existing ABM literature is exploring liquidity in an OTC government bond market to understand how diverse factors contribute to stability or fragility. A growing body of literature is examining the efficiency of markets and the potential for "chaos" to be a distinct market feature \cite{BROCK19981235} \cite{Ingladaperez2020}. Adding to the complexity of market microstructure analysis. 

\section{Model: A Bespoke ABM Implementation Based on Sugarscape}

We build a bespoke ABM, drawing heavily from Sugarscape, the model introduced by \cite{Axtell} used to simulate complex, artificial societies through agent-environment interactions. Many adaptations of this work exist \cite{Pike_2019, Masad_2015, Pike_2021, Kazil_2020, Axtell2022}.  The model's ability to represent small populations of heterogeneous agents, interacting asynchronously, aligns closely with the bilateral structure of government bond markets, making it an ideal foundation for the bespoke ABM used in this research.  Other ABM model frameworks were considered (such as \cite{ABIDES}), however these tend to require extensive calibration and do not suit the bilateral market focus of this study. To develop our bespoke model, several open-source tools were leveraged, including repositories from GitHub \cite{python-mesa-2020}, \cite{SpecofSugar}  following the framework provided by \cite{Pike_2019}.

\subsection{Agents}
The model consists of \(N\) agents, representing market makers (MM's).  We will use the term agent and MM interchangeable in this paper.  In the Australian market, four key institutional market makers dominate, reflecting a structure where fewer MMs interact with a large number of clients, as noted by \cite{Cheshire2015}. Additionally we model a cost structure for agents to represent fixed operational costs incurred by agents per time step. 

\subsubsection{Vision: Client base and breadth}
All financial firms have clients, and in the case of MM's these are the firms and people they are required to buy and sell bonds with under the terms of their government registration.  Within the model we characterise the difference in MM size through their different client breadth (i.e. grid vision range) \(v_i\), determining how far they can access client resources across the landscape. For an agent at position \( (x_i, y_i) \), the set of visible cells is: \( \mathcal{V}_i = \{ (x, y) \mid |x - x_i| \leq v_i \text{ and } |y - y_i| \leq v_i \} \).
This range is randomly assigned as \(v_i \sim \text{Uniform}(1, 49)\) such that the maximum any MM can access is the whole of the grid (i.e. less than 50 units) and remains constant during the simulation.

\subsubsection{Cost of Doing Business: Metabolism}
Each MM incurs a cost at each time step, representing various business overheads, modelled as a form of metabolism. The bond and cash holdings of an agent \(i\) at each time step are updated as follows: \(A_b(t+1) = A_b(t) + (b_r - m_b)
\), \(A_c(t+1) = A_c(t) + (c_r - m_c) \).  Where \(m_b\) and \(m_c\) represent the bond and cash metabolic rates. These costs are assigned at the start and remain constant, influencing an agent's ability to survive and stay in business.

\subsection{Clients}
Clients are represented by static locations a \( 50 x50\) grid, totaling 2,500 distinct clients based on data from \cite{Pinter2023}. Each grid cell \((x, y)\) corresponds to a client holding bonds \(B_{(x, y)}\) and cash \(C_{(x, y)}\), modelled as a tuple: \( C_{i} = R_{(x, y)} = (B_{(x, y)}, C_{(x, y)})\).  Each client is assigned a heterogeneous quantity of a generic government bond, and cash amount. Bonds and cash are distributed in mounds across the grid, with values decreasing from the centre of the mounds outward upon the grid. Due to the lack of observable real-world distribution of bondholders, we chose to concentrate resources in four distinct regions on the grid. Future research may refine this assumption as more data becomes available.


\subsection{MM servicing clients}
Market makers (MMs) are assigned a fixed client range, modeled as a form of grid "vision," allowing them to "see" client locations within a maximum grid distance at each stage. This enables MMs to locate client resources within their field of vision on the grid. Agents exit the market when either their bond or cash holdings reach zero (\(A_b = 0\) or \(A_c = 0\)).  MMs acquire bonds and cash from passive clients, reflecting the key aspect of over-the-counter (OTC) markets where all client trading involves a market maker—a feature effectively captured in our ABM design.

\subsection{Market Maker Trading}
After servicing clients at each time step, market makers (MMs) consider potential trades with other agents within their field of vision, provided the trade improves welfare for both parties. Welfare improvement addresses imbalances in cash or bonds held by an MM, aligning with rational and legal trading behaviour as outlined in finance literature and regulations \cite{BOEOpps}. Each agent’s welfare for bonds and cash is calculated as: \( Welfare_{a_{i},b} = A_{b}^\frac{m_{b}}{m_{b} + m_{c}}, \quad Welfare_{a_{i},c} = A_{c}^\frac{m_{c}}{m_{b} + m_{c}} \). 

Where \(A_b\) and \(A_c\) are the agent's accumulations of bonds and cash, and \(m_b\) and \(m_c\) are the respective costs.  Small cost values relative to initial accumulations are used.

For a trade to occur, agents compare the product of their current welfare to the potential new welfare after the trade: \[ \hat{Welfare_{a_{i},b}} \times \hat{Welfare_{a_{i},c}} \geq {Welfare_{a_{i},b}} \times {Welfare_{a_{i},c}}
\]
This condition must hold for both trading agents, ensuring mutual welfare improvement. This eliminates the need for a price mechanism, as agents trade based on relative needs, reflecting the limited price transparency in the modelled markets.


Agents compare their Marginal Rate of Substitution (MRS), indicating their relative need for bonds or cash. Trades occur when agents have offsetting needs, and the exchange quantity is determined using the geometric mean of their MRS, reducing bias from extreme values as proposed by \cite{RichGmean}.

The MRS for an agent \( a_i \) is defined as \(
\text{MRS\(a_i\)} = \frac{\left(\frac{A_{c}}{m_{c}}\right)}{\left(\frac{A_{b}}{m_{b}}\right)}
\).  A higher MRS reflects a stronger preference for cash over bonds, and vice versa, based on survival needs in the model.

\begin{definition}
\textbf{We define an ABM formally as:}
\( \mathcal{M} = \left\langle N, \mathcal{L}, \mathcal{S}, \mathcal{P}, \mathcal{D}, \mathcal{A}, f, T \right\rangle \), where \( N = \{a_1, a_2, \dots, a_n\} \) is the set of market making agents, with \( n = 4 \) in this case. \( \mathcal{L} \) represents the set of landscape states, each containing assets  \( R = \{b_r, c_r\} \), corresponding to bond and cash quantities of individual clients. Each agent \( a_i \) has a state \( \mathcal{S}_i \), which includes bond and cash usage (i.e. their individual cost structure) rates \( m_b, m_c \), accumulations \( A_b, A_c \), and a client base breadth \( v_i \). The perception functions \( \mathcal{P}_i \) map the landscape and other agents' states to a perceived state, \( p_{ij}: \mathcal{L} \cup \mathcal{S} \rightarrow \mathbb{R} \), while the decision rules \( \mathcal{D}_i \) map perceptions and states to actions, \( d_{ik}: \mathcal{P}_i \times \mathcal{S}_i \rightarrow \mathcal{A}_i \). Agent actions \( \mathcal{A}_i \) include client servicing, trading with other market makers, and ceasing operations. The landscape evolves according to the transition function \( f: \mathcal{L} \times \mathcal{A} \rightarrow \mathcal{L} \), with no bond or cash replenishment, and the model progresses through discrete time steps, \( T = \mathbb{N} \).

\end{definition}

\subsection{Model Dynamics}
The model dynamics are governed by the iterative interactions of the components in \( \mathcal{M} \). At \( t = 0 \), MMs are randomly assigned positions on a grid \( X \times Y \), with no preferential locations, simulating the varying client access that MM's experience daily. Each MM perceives the environment \( \mathcal{L} \) and the states of other agents \( \mathcal{S} \) within their field of vision, then selects an action based on its decision rule. MMs' accumulated bonds and cash are decremented to cover business costs, set at the inception of each epoch. MMs also have the option to trade with each other when a utility difference exists, exchanging bonds or cash based on perceived liquidity imbalances between their holdings.  If an MM's resources are depleted (\(A_b = 0\) or \(A_c = 0\)), they cease operations.

After each action, the landscape and agent states are updated, adjusting resource variables \(b_r\) and \(c_r\) accordingly. The process iterates across all agents and over time until all agents have ceased operations or the maximum time \( T \) is reached.  The ABM does not allow any agent coalition formation, in light with real world regulatory constraints on market makers.  

Our model makes key assumptions to streamline its design. First, clients are passive, always ready to trade, and fully transparent about their resources within their client base. While this simplifies interactions and limits realism, it allows the focus to remain on agent-to-agent interactions, justified by the lack of available data on client holdings. Second, we assume that market makers (MMs) truthfully reveal their resource levels to others and do not store data about previous accumulations, aligning with market regulations on fair dealing. We do not model MM strategic coalition formation, which can be contrary to policy \cite{tpicap2021best}. These assumptions simplify the model by avoiding complex information exchange dynamics and fit within the regulated bond trading environment.

\subsection{Available Data}
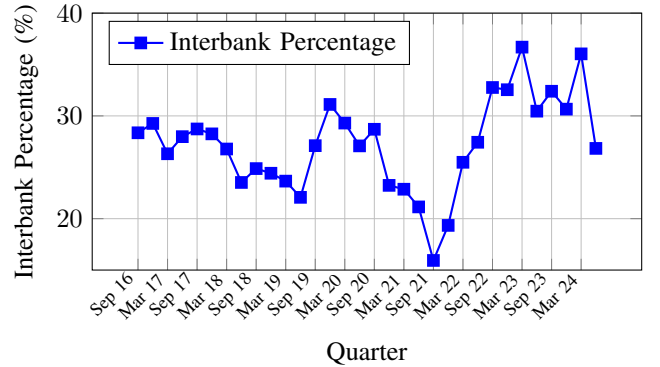
\begin{figure}[htpb]
\centering
\begin{tikzpicture}
\begin{axis}[
    width=0.49 \textwidth,
    height= 5cm,
    xlabel={Quarter},
    ylabel={Interbank Percentage (\%)},
    xticklabel style={
        font=\scriptsize,
        rotate=45,
        anchor=east
    },
    symbolic x coords={
        Sep 16, Dec 16, Mar 17, Jun 17, Sep 17, Dec 17, Mar 18, Jun 18, Sep 18, Dec 18,
        Mar 19, Jun 19, Sep 19, Dec 19, Mar 20, Jun 20, Sep 20, Dec 20, Mar 21, Jun 21,
        Sep 21, Dec 21, Mar 22, Jun 22, Sep 22, Dec 22, Mar 23, Jun 23, Sep 23, Dec 23,
        Mar 24, Jun 24
    },
    xtick={
        Sep 16, Mar 17, Sep 17, Mar 18, Sep 18, Mar 19, Sep 19, Mar 20, Sep 20, Mar 21,
        Sep 21, Mar 22, Sep 22, Mar 23, Sep 23, Mar 24
    },
    ymin=15, ymax=40,
    grid=both,
    grid style={line width=.1pt, draw=gray!20},
    major grid style={line width=.2pt, draw=gray!50},
    tick align=inside,
    ymajorgrids=true,
    xmajorgrids=true,
    legend pos=north west,
    scaled x ticks=false,
]

\addplot[
    color=blue,
    mark=square*,
    thick,
]
coordinates {
    (Sep 16,28.357)
    (Dec 16,29.266)
    (Mar 17,26.316)
    (Jun 17,27.980)
    (Sep 17,28.729)
    (Dec 17,28.252)
    (Mar 18,26.779)
    (Jun 18,23.524)
    (Sep 18,24.873)
    (Dec 18,24.420)
    (Mar 19,23.654)
    (Jun 19,22.075)
    (Sep 19,27.104)
    (Dec 19,31.115)
    (Mar 20,29.303)
    (Jun 20,27.083)
    (Sep 20,28.698)
    (Dec 20,23.240)
    (Mar 21,22.865)
    (Jun 21,21.144)
    (Sep 21,15.935)
    (Dec 21,19.344)
    (Mar 22,25.486)
    (Jun 22,27.435)
    (Sep 22,32.764)
    (Dec 22,32.547)
    (Mar 23,36.691)
    (Jun 23,30.461)
    (Sep 23,32.398)
    (Dec 23,30.656)
    (Mar 24,36.036)
    (Jun 24,26.837)
};

\addlegendentry{Interbank Percentage}

\end{axis}
\end{tikzpicture}
\caption{Quarterly Interbank Percentage of Secondary Australian Government Bond Trading}
\label{fig:interbank_percentage}
\end{figure}

Publicly available data from the AOFM reports the amount of MM to MM trading as a share of total secondary market trading in Australian government bonds.  This data is released quarterly, with data available at a quarterly level from September 2016 to June 2024, aggregated across all market participants.  See Table \ref{tab:Summary} and Figure \ref{fig:interbank_percentage}.

\begin{table}[h]
\centering
\begin{tabular}{lr}
\hline
\textbf{Statistic} & \textbf{Value (\%)} \\
\hline
Mean & 27.23 \\
Median & 27.27 \\
Quarterly Std Dev & 4.54 \\
Minimum & 15.94 \\
Maximum & 36.69 \\
First Quartile (Q1) & 23.85 \\
Third Quartile (Q3) & 30.17 \\
Interquartile Range (IQR) & 6.33 \\
\hline
\end{tabular}
\caption{Aggregated quarterly Summary Statistics of Interbank trading volume market share: Sep 2016 to June 2024}
\label{tab:Summary}
\end{table}


\section{Results}
We calibrated our ABM to the Australian MM environment, successfully replicating the observed first and second order attributes. Using the design and model variables listed in Table \ref{tab:Golidlocks}, we conducted 100 simulations (epochs), producing over 200,000 agent interactions. We explored calibration sets opportunistically rather than exhaustively testing all permutations of model variables, which would be intractable. We refer to this configuration as the "Australian calibration" or \textbf{(Hypothesis 1: HP1)}.  

Our simulations reveal a median MM-to-MM trading occurrence of 28.74\% of all interactions. This median trading activity aligns closely with data published by the Australian Office of Financial Management (AOFM), which reports an eight-year average interbank bond turnover of 27.3\% of traded volumes. This close correspondence validates our model's ability to replicate real-world trading patterns.

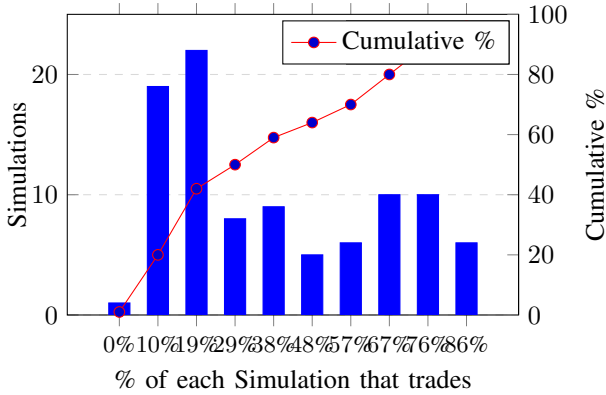
\begin{figure}[htpb]
\centering
\begin{tikzpicture}
\def\plotwidth{6cm}
\def\plotheight{4cm}
\begin{axis}[
    ybar,
    bar width= 8pt,
    width=\plotwidth,         
    height=\plotheight,       
    xlabel={\% of each Simulation that trades},
    ylabel={Simulations},
    ymin=0,
    ymax=25,
    xtick=data,
    symbolic x coords={0,10,19,29,38,48,57,67,76,86},
    xticklabel={\pgfmathprintnumber{\tick}\%},
    enlarge x limits=0.15,
    ylabel near ticks,
    ymajorgrids=true,
    grid style={dashed,gray!30},
    scale only axis,
    ytick pos=left,            
    yticklabel pos=left,       
    xticklabel style={font=\small},  
    ylabel style={at={(axis description cs:-0.07,.5)},anchor=south}
]
\addplot+[ybar, fill=blue] coordinates {
    (0,1)
    (10,19)
    (19,22)
    (29,8)
    (38,9)
    (48,5)
    (57,6)
    (67,10)
    (76,10)
    (86,6)
};
\addlegendentry{Frequency}
\end{axis}
\begin{axis}[
    width=\plotwidth,         
    height=\plotheight,       
    ymin=0,
    ymax=100,
    axis y line*=right,      
    xlabel={\% of each Simulation that trades},
    ylabel={Cumulative \%},
    ylabel near ticks,
    xtick=data,
    symbolic x coords={0,10,19,29,38,48,57,67,76,86},
    xticklabel={\pgfmathprintnumber{\tick}\%},
    enlarge x limits=0.15,    
    yticklabel pos=right,     
    ytick pos=right,          
    axis x line=none,         
    ymajorgrids=false,
    scale only axis,
    hide x axis              
]
\addplot+[mark=*, color=red] coordinates {
    (0,1.00)
    (10,20.00)
    (19,42.00)
    (29,50.00)
    (38,59.00)
    (48,64.00)
    (57,70.00)
    (67,80.00)
    (76,90.00)
    (86,96.00)
};
\addlegendentry{Cumulative \%}
\end{axis}
\end{tikzpicture}
\caption{A=4 Distribution across Runs by Trading Percent} 
\label{fig:A4}
\end{figure}

The results demonstrate a high degree of variance across simulations, as illustrated in Figure \ref{fig:A4}. This variance is expected given the large degrees of freedom in the initial calibrations for each simulation. On average, the trading frequency in the model aligns closely with reported aggregate trading volumes, serving as a reasonable approximation for such an opaque market. Further research is needed to determine whether these factors contribute to the market "chaos" 
Additionally, we note that the reported figures are aggregated over three months of daily trading without reporting variance or week-to-week volatility. Nevertheless, the range of trading interactions remains broad, with summary statistics provided in Table \ref{tab:epoch_trade_summary_stats} (rounded to two decimal places).

\begin{table}[h!]
\centering
\begin{tabular}{|c|c|}
\hline
\textbf{Variable} & \textbf{Description} \\ \hline
Number of Agents & 4 \\ \hline
Client Breadth & 1 to 50 units (random) \\ \hline
Cost & 1 to 5 units per time step (random) \\ \hline
Initial Resource Accumulation & 35 to 55 units (random) \\ \hline
Size of Client Base & 2500 clients \\ \hline
Trading Occurrence & Median 28.74\% per model simulation \\ \hline
Number of Simulations & Over 100 \\ \hline
Time Steps per Simulation & Up to 1500 \\ \hline
Agent Moves per Trading Hour & Approximately 200 \\ \hline
Simulation Order & Asynchronous and randomised \\ \hline
\end{tabular}
\caption{Summary of Variables in the Simulation}
\label{tab:Golidlocks}
\end{table}

\begin{table}[ht!]
\centering
\begin{tabular}{|l|c|}
\hline
\textbf{Statistic} & \textbf{Epoch Trade Percentage} \\ \hline
\textbf{Median}      & 28.54                           \\ \hline
\textbf{Std Dev}   & 27.48                            \\ \hline
\textbf{Min}       & 1.22                             \\ \hline
\textbf{25\%}      & 11.03                           \\ \hline
\textbf{50\%}      & 28.54                            \\ \hline
\textbf{75\%}      & 59.77                            \\ \hline
\textbf{Max}       & 95.11                            \\ \hline
\end{tabular}
\caption{Summary Statistics for Epoch Trade Percentage}
\label{tab:epoch_trade_summary_stats}
\end{table}

\subsection{Market Makers Numbers: Do we need more?}
We explore the hypothesis that agent diversity, rather than simply increasing the number of agents, has a positive impact on interactions and trading liquidity. To illustrate our findings, we report two examples, each consisting of 100 simulation epochs of agent models.

\begin{itemize}
    \item \textbf{(Hypothesis 2: HP2)} - We keep \(N = 4\) and maintain the same costs but reduce the possible range of client base sizes from a theoretical maximum of 50 units to just 5 units. In each epoch, agents are randomly assigned a value within this reduced range for the duration of the simulation. By narrowing the range from 1–50 units to 1–5 units, we significantly decrease the maximum client base size and reduce agent-to-agent diversity in client base sizes (as agents can now assume only one of 5 possible client size values instead of 50). The result is a series of simulations where almost no MM-to-MM trading occurs. In fact, over 73\% of epochs exhibit no market maker trading whatsoever (see figure \ref{fig:A4_M5V5}). This contrasts with the Australian calibration, where agents with client ranges of 1–50 units engage in MM-to-MM trading on average 28\% of the time, and 100\% of epochs display some market maker trading, however small.

\begin{figure}[htpb]
\centering
\begin{tikzpicture}
\def\plotwidth{6cm}
\def\plotheight{4cm}
\begin{axis}[
    ybar,
    bar width= 9pt,
    width=\plotwidth,         
    height=\plotheight,       
    xlabel={\% of each Simulation that trades},
    ylabel={Simulations},
    ymin=0,
    ymax=150,
    xtick=data,
    symbolic x coords={0,1,2,3,4,5,6,7,8,9,10},
    xticklabel={\pgfmathprintnumber{\tick}\%},
    enlarge x limits=0.08,
    ylabel near ticks,
    ymajorgrids=true,
    grid style={dashed,gray!30},
    scale only axis,
    ytick pos=left,            
    yticklabel pos=left       
]
\addplot+[ybar, fill=blue] coordinates {
    (0,147)
    (1,20)
    (2,9)
    (3,2)
    (4,6)
    (5,4)
    (6,0)
    (7,4)
    (8,1)
    (9,3)
    (10,3)
};
\addlegendentry{Frequency}
\end{axis}
\begin{axis}[
    width=\plotwidth,         
    height=\plotheight,       
    ymin=0,
    ymax=100,
    axis y line*=right,      
    xlabel={\% of each Simulation that trades},
    ylabel={Cumulative \%},
    ylabel near ticks,
    xtick=data,
    symbolic x coords={0,1,2,3,4,5,6,7,8,9,10},
    xticklabel={\pgfmathprintnumber{\tick}\%},
    enlarge x limits=0.08,
    yticklabel pos=right,     
    ytick pos=right,          
    axis x line=none,         
    ymajorgrids=false,
    scale only axis,
    hide x axis              
]
\addplot+[mark=*, color=red] coordinates {
    (0,73.0)
    (1,84.0)
    (2,88.0)
    (3,89.0)
    (4,92.0)
    (5,94.0)
    (6,94.0)
    (7,96.0)
    (8,96.5)
    (9,98.0)
    (10,99.5)
};
\addlegendentry{Cumulative \%}
\end{axis}
\end{tikzpicture}
\caption{A=4  Client and Cost reduction (median 0.8\%). Stressing the Australian calibration to much lower client breadth and business cost effectively shuts down market maker trading to \textbf{zero} in 73\% of epochs with an average trade amount of less than 1\% of all activity}
\label{fig:A4_M5V5}
\end{figure}
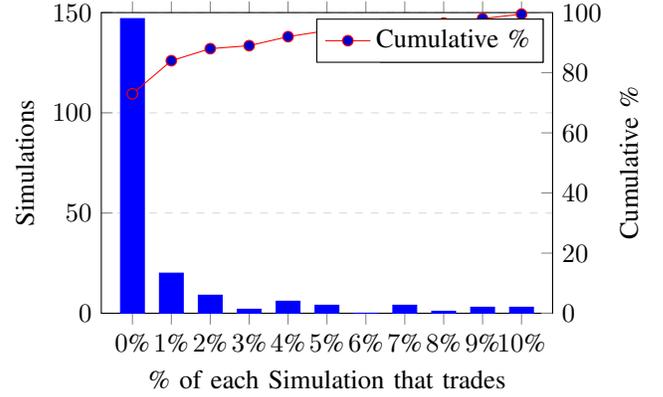


    \item \textbf{(Hypothesis 3: HP3)} To test the impact of number of clients, We keep the same reduced client base and same business costs as in \textbf{HP2} but increase agents to 16 (or 4 times as many). We see more trades than in \textbf{HP2}, but still far fewer MM to MM trades than the Australian calibration set - just 6.418\% of the time (see figure \ref{fig:A16}).

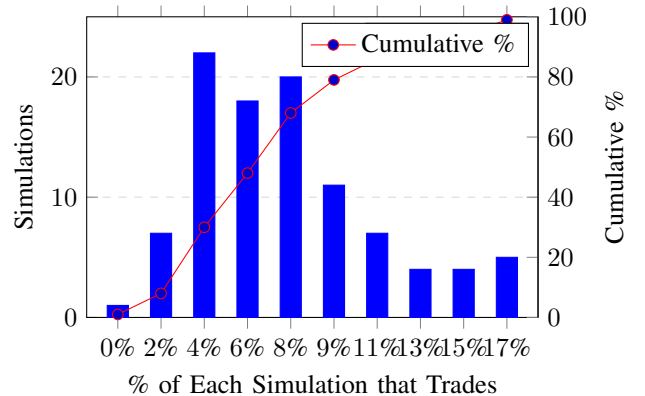
\begin{figure}[htpb]
\centering
\begin{tikzpicture}
\def\plotwidth{6cm}
\def\plotheight{4cm}
\begin{axis}[
   ybar,
   bar width=8pt,
   width=\plotwidth,         
   height=\plotheight,       
   xlabel={\% of Each Simulation that Trades},
   ylabel={Simulations},
   ymin=0,
   ymax=25,
   xtick=data,
   symbolic x coords={0,2,4,6,8,9,11,13,15,17},
   xticklabel={\pgfmathprintnumber{\tick}\%},
   enlarge x limits=0.08,
   ylabel near ticks,
   ymajorgrids=true,
   grid style={dashed,gray!30},
   scale only axis,
   ytick pos=left,            
   yticklabel pos=left,       
]
\addplot+[ybar, fill=blue] coordinates {
   (0,1)
   (2,7)
   (4,22)
   (6,18)
   (8,20)
   (9,11)
   (11,7)
   (13,4)
   (15,4)
   (17,5)
};
\addlegendentry{Frequency}
\end{axis}
\begin{axis}[
   width=\plotwidth,         
   height=\plotheight,       
   ymin=0,
   ymax=100,
   axis y line*=right,       
   ylabel near ticks,
   ylabel={Cumulative \%},
   xtick=data,
   symbolic x coords={0,2,4,6,8,9,11,13,15,17},
   xticklabel={\pgfmathprintnumber{\tick}\%},
   enlarge x limits=0.08,
   yticklabel pos=right,     
   ytick pos=right,          
   axis x line=none,         
   ymajorgrids=false,
   scale only axis,
   hide x axis              
]
\addplot+[mark=*, color=red] coordinates {
   (0,1.00)
   (2,8.00)
   (4,30.00)
   (6,48.00)
   (8,68.00)
   (9,79.00)
   (11,86.00)
   (13,90.00)
   (15,94.00)
   (17,99.00)
};
\addlegendentry{Cumulative \%}
\end{axis}
\end{tikzpicture}
\caption{$N = 16$ - distribution of trade occurrences, as a percentage for each simulation. Cumulative distribution shows a mean of just 6.49\% as reported}
\label{fig:A16}
\end{figure}


\end{itemize}
Whilst not a complete analysis of all possibilities, these two specific examples demonstrate that the number of agents cannot make up for the benefits that come from having a wide variety of agent client sizes, such as in \textbf{HP1}. We document that by increasing agents, while reducing client size and keeping cost levels (and range) low, dramatically reduce trading in all epochs.  As agents select a level from a lower range of client options, reducing client breath has the impact of decreasing agent heterogeneity also.  Market makers in these scenarios simply have less differentiation between themselves also. Even when increasing the number of agents to 16 (significantly beyond that of our Australian market calibration of just 4), we see that the amount of trading between agents is significantly reduced from 28\% in HP1 to under 7\% in HP3 and virtually halted in HP2. From this we conclude that increasing the heterogeneity of market makers rather than simply the number of agents contributes to greater market trading activity and hence liquidity.


\subsection{Impact of lowering Agent costs and Market Stability}

The concept of agent lifespan, in terms of time steps, was evaluated in relation to stability and liquidity provision. The results demonstrate that low agent costs tend to enhance agent longevity, which can be associated with higher stability in market trading. We utilise the two tests again, this time looking at populations of 4 agents across 100 epochs where client breadth is set to be diverse (1 - 50 units in line with HP1) but costs are set to be higher (with a range of 5-10 units).  This test can then be directly compared to the tests using the Australian MM calibration simulations, which has the same calibration set but a cost range of just 1- 5 units.

\begin{itemize}

    \item \textbf{HP1} the Australian market calibration, approximately 27.2\% of interactions are between market makers. Also, 68\% of epochs had at least one agent still servicing clients and "alive" at the maximum time step of \(t = 1500\).  We can see in figure\ref{fig:A4_bondCost} and figure\ref{fig:A4_CashCost} that agents in epochs where they had lower costs have longer life spans. 

\begin{figure}[htbp]
    \centering
    \begin{subfigure}{.49\columnwidth}  
        \includegraphics[width=\linewidth]{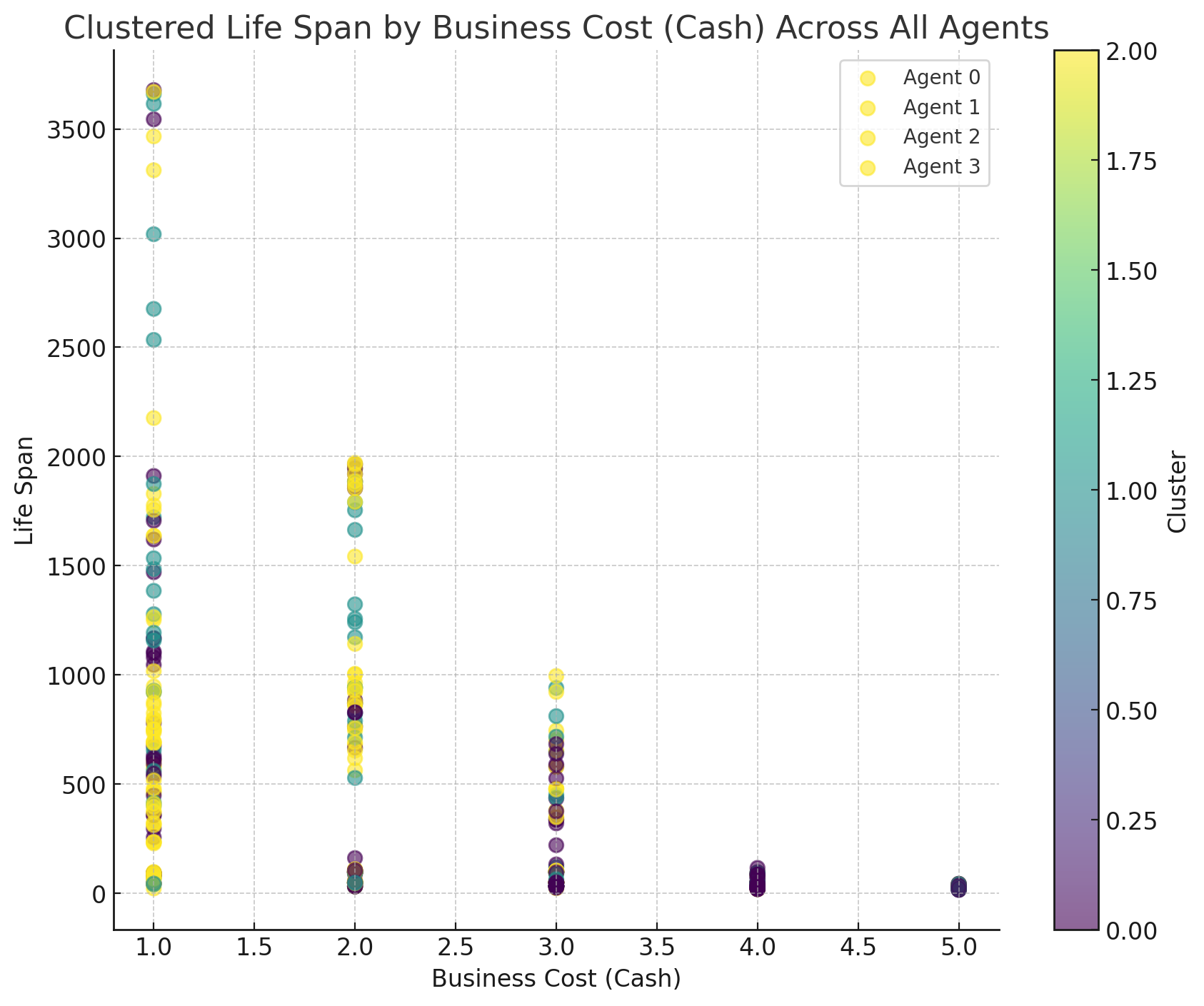}
        \caption{Impact of cash costs}
        \label{fig:A4_CashCost}
    \end{subfigure}%
    \begin{subfigure}{.49\columnwidth}  
        \includegraphics[width=\linewidth]{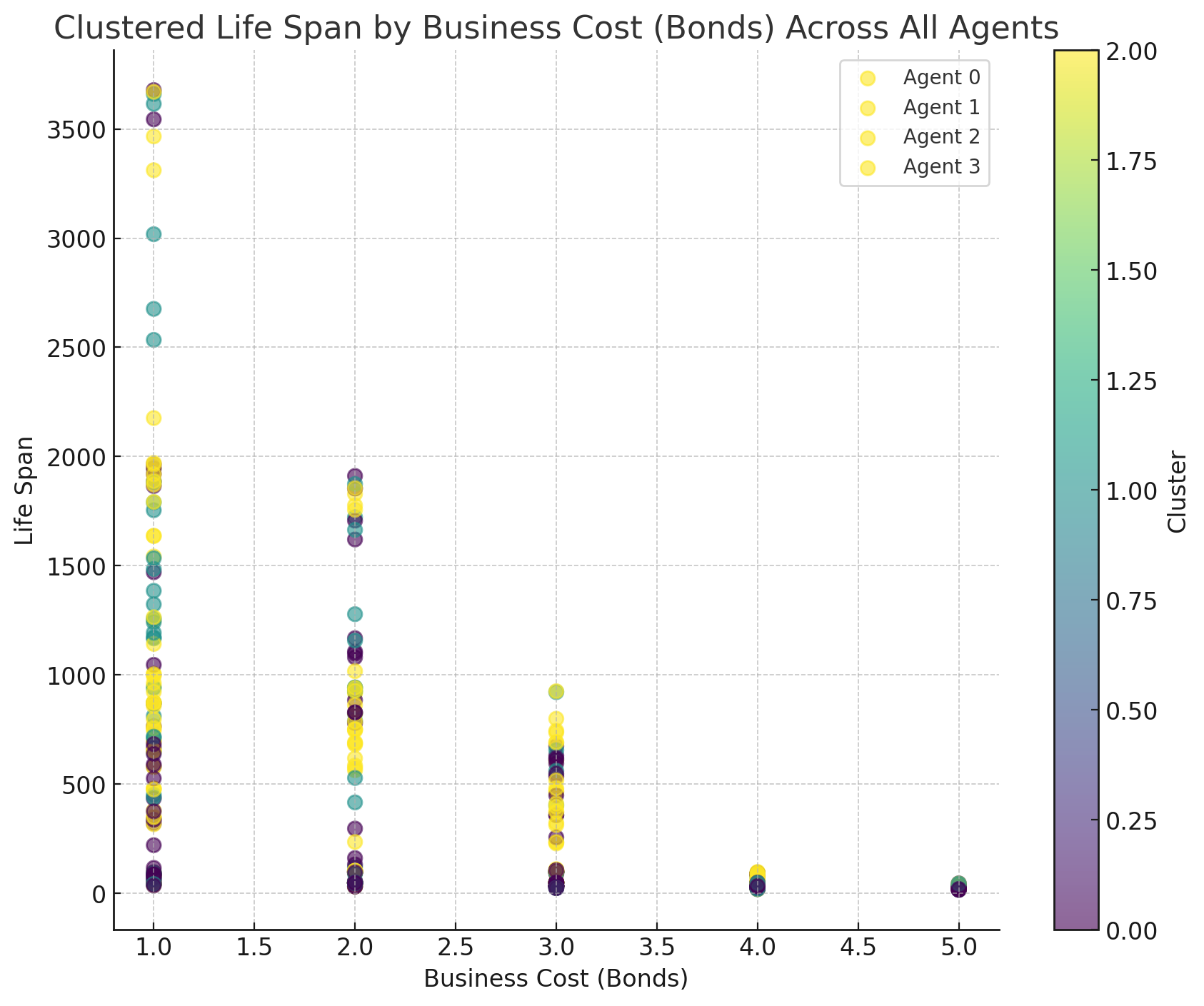}
        \caption{Impact of bond costs}
        \label{fig:A4_bondCost}
    \end{subfigure}
\end{figure}

\item \textbf{(Hypothesis 4: HP4)} - We stress costs to double the level in \textbf{HP1}. The results are quite stark: average life spans of each epoch is less 17 time steps, with less than 4,000 interactions across all 100 epochs. Individual epochs show significant early collapse (around 10 or so time steps), but, across a small number of time steps, there is a large degree of trading with other market makers (median of 46\%).  The key outcome appears to be that increasing (doubling) costs significantly weakens the trading landscape (see figure\ref{fig:A14_high}) leading to situations where the landscape fails to function, meeting the definition of instability provided by the World Bank \cite{WorkBank24}.

\begin{figure}   
                \centering
                \includegraphics[width=0.95\linewidth]{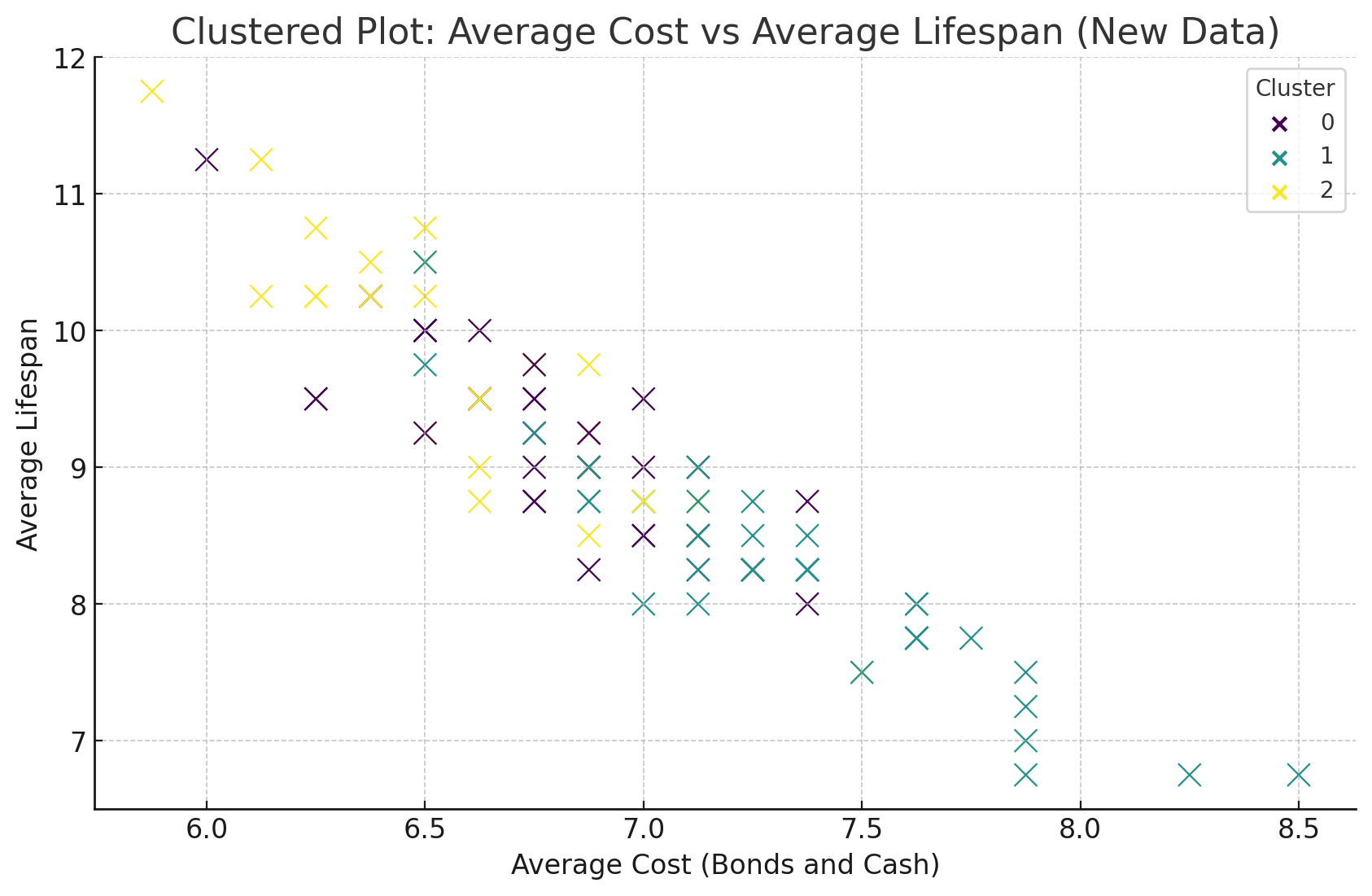}
                \caption{High costs, very short life span: Market Instability}
                \label{fig:A14_high}
            \end{figure}
    
\end{itemize}

\section{Summary, Conclusions and Future Work}

Our work contributes to the literature by offering a simulation framework that captures key elements of OTC bond market structure, while also accounting for liquidity and market stability concerns. By leveraging a bespoke ABM, we replicate observable market phenomena in the Australian government bond market. This allows us to explore theoretical scenarios that can inform policy decisions, such as the impact of market maker client diversity and cost structures.

Maintaining stable market functioning is not only vital for societal functions, but is also essential for regulatory bodies, such as the World Bank, International Monetary Fund and Bank of England. Balancing client servicing and agent-to-agent trade is fundamental to ensure the network's viability. We show that MM trading can be effectively halted in 73\% of simulations (\textbf{HP2}) when client ranges are reduced to a maximum of 10\% of the Australian market calibration. However, increasing the number of agents in this scenario from 4 to 16 (\textbf{HP3}) improves liquidity but it remains significantly less activity and liquidity than \textbf{HP1} - the Australian market calibration.

Our final conjecture tests the impact of trading costs on stability looking at MM trading and lifespan. We see in \textbf{HP4}, that doubling trading costs from the simulations in \textbf{HP1}, produce drastically smaller lifespans of agents (HP1 over 68\% of agents lived past 1500 time steps, in HP4, no agent lived beyond just \textbf{17} time steps. From this we conclude the markets with lower costs find stability easier to maintain. 

Future work may explore enhancing the model's richness and accuracy by incorporating more comprehensive market data as it becomes available. Additionally, we intend to explore alternative approaches to model client behaviours and asset distributions to further enrich our model's ability to reflect real-world market behaviour.  Additional analysis of market internal mechanisms will also be carried out.

    \bibliographystyle{IEEEtran}

    \bibliography{Main_mybibliography}

\end{document}